\documentclass[aps,prl,twocolumn,superscriptaddress,showpacs]{revtex4}
\usepackage{graphicx}
\usepackage{amsmath}
\newcommand{\hmo}{HoMnO$_3$}

\newcommand{\mb}{\mathbf}
\newcommand{\nn}{\nonumber}

\begin{document}

\title{Dual Nature of Improper Ferroelectricity in a Magnetoelectric Multiferroic}
\author{S. Picozzi}
\affiliation{Consiglio Nazionale delle Ricerche - Istituto Nazionale di Fisica della Materia (CNR-INFM),
CASTI Regional Lab., 67010 L'Aquila, Italy}
\author{K. Yamauchi}
\affiliation{Consiglio Nazionale delle Ricerche - Istituto Nazionale di Fisica della Materia (CNR-INFM),
CASTI Regional Lab., 67010 L'Aquila, Italy}
\author{B. Sanyal}
\affiliation{Theoretical Magnetism Group, Department of Physics, Uppsala University, Box-530, SE-75121, Sweden}
\author{I. A. Sergienko}
\affiliation{Materials Science \& Technology Division, Oak Ridge
National Laboratory, Oak Ridge, TN 37831, USA}
\affiliation{Dept. of Physics \& Astronomy, The University of Tennessee,
Knoxville, TN 37996, USA}
\author{E. Dagotto}
\affiliation{Materials Science \& Technology Division, Oak Ridge
National Laboratory, Oak Ridge, TN 37831, USA}
\affiliation{Dept. of Physics \& Astronomy, The University of Tennessee,
Knoxville, TN 37996, USA}
\pacs{75.80.+q, 75.50.Ee,  77.80.-e, 75.47.Lx}

\begin{abstract}
Using first principles calculations, we study the microscopic origin of
ferroelectricity (FE) induced by magnetic order in the orthorhombic \hmo. 
We obtain the largest ferroelectric polarization observed in the
whole class of improper magnetic ferroelectrics to date.
We find that the two proposed mechanisms for FE in multiferroics, 
lattice- and electronic-based, are simultaneously active in this compound:
a large portion of the ferroelectric polarization arises
due to quantum-mechanical effects of electron orbital polarization, 
in addition to the conventional polar atomic displacements. An interesting 
mechanism for switching the magnetoelectric domains by an electric field via a
180$^{\circ}$ coherent rotation of Mn spins is also proposed.
\end{abstract}

\maketitle 

Magnetoelectric materials owing their improper ferroelectric (FE) order
to symmetry breaking magnetic structures are drawing enormous recent
interest~\cite{kimura,Hur04,review1}. One of the 
fundamental problems in this area is the understanding of 
the microscopic origin of their electric polarization.
Two basic mechanisms have been proposed in model studies. 
According to one of them, magnetic ordering results in the hybridization 
of electronic orbitals producing a polar charge distribution~\cite{nagaosa,Jia07}.
The other, more conventional approach, views the displacements of 
ions from their centrosymmetric positions as the primary source of the
polarization~\cite{tbmno32,Sergienko06,ivan}.
Extensive experimental studies have not been able to distinguish
between the two possibilities due to very small values
of the polarization $P$ found in this class of multiferroics, such as $P < 0.1\, \mu$C/cm$^2$ in
TbMnO$_3$ and TbMn$_2$O$_5$~\cite{kimura, Hur04}.

\begin{figure}[t] 
\resizebox{78mm}{!}{\includegraphics{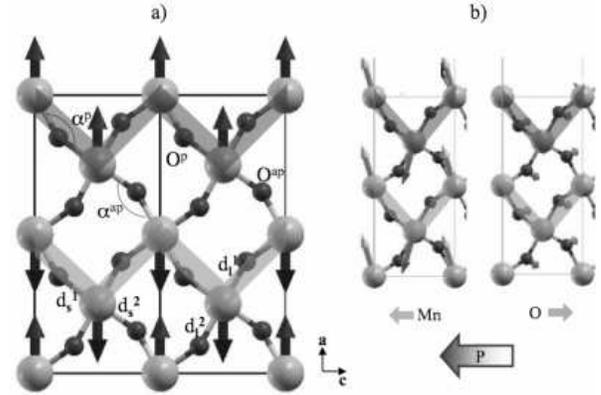}}
\caption{(Color online)
a) The in-plane arrangement of Mn and O atoms. Arrows denote
  the direction of spins and antiferromagnetically-coupled zig-zag
  spin-chains are highlighted by pink and blue shaded
  areas. Structural parameters reported in Table~\ref{tabella} are
  shown. 
b) Arrows show the directions of the in-plane ionic displacements for
  Mn (left) and O (right) in the AFM-E phase. The thick arrows at the
  bottom show the direction of the resulting displacements the Mn and
  O sublattices and $\mb P$.} 
\label{fig:struc}
\end{figure}

In a quest for higher $P$, a recent model Hamiltonian study concentrated on 
the collinear antiferromagnetic-E (AFM-E) spin configuration, where
ferromagnetic zigzag spin-chains in the MnO$_2$ planes are
antiferromagnetically coupled with respect to both adjacent in-plane
chains (see Fig.~\ref{fig:struc}a) and out-of-plane stacked chains, 
as found in the orthorhombic \hmo\ and other perovskite compounds~\cite{Munoz01, ivan}.
The predicted polarization $P=0.5-12\, \mu$C/cm$^2$ was much higher than in 
other improper magnetic ferroelectrics. However, pyroelectric current
measurements on bulk polycrystalline samples revealed ferroelectricity
in \hmo\ with $P$ of only less than $2$ nC/cm$^2$~\cite{Lorenz06}. 

First-principles calculations appear to be well-suited to address the
two issues discussed above: (1) they can reveal the dominant mechanism 
of improper FE polarization in magnetically ordered compounds, since the electronic 
structure and lattice distortions can be treated simultaneously, and (2) they
can solve the theory-experiment discrepancy for \hmo in the AFM-E phase 
by clarifying the true value of the electric polarization.
We perform simulations based on the generalized gradient
approximation (GGA)~\cite{pbe} to density functional (DF) theory using the
Vienna Ab-initio Simulation Package (VASP)~\cite{vasp} and the projector-augmented-wave
pseudopotentials~\cite{paw}. Ho 4$f$ electrons were assumed as frozen
in the core. The plane wave energy cut-off was set to 500 (400) eV for
the collinear (non-collinear) calculations. The Brillouin zone
sampling was performed using the 3x4x6 shell~\cite{monkh}. The GGA+U~\cite{anisimov}
calculations are performed by applying a Hubbard-like potential for
Mn $d$ states, with U ranging from 0 to 8 eV; accordingly, we chose J = 0.15~U. The Berry phase
approach~\cite{berry1,berry2} was used to calculate $P$,
integrating over six {\bf k}-point strings parallel to the $c$ axis,
each string containing 6 {\bf k}-points. Non-collinear calculations
were performed according to Ref.~\cite{hobbs}. Spin-orbit coupling (SOC) was neglected. 
As for the structural details, we chose the experimental lattice
constants~\cite{hmo1} for \hmo\  for the orthorhombic unit-cell (space group $Pnma$, $a$ =
5.835 \AA, $b$ = 7.361 \AA\ and $c$ = 5.257 \AA) and performed atomic
relaxations  until the Hellman-Feynman forces were below 0.015
eV/\AA.

\begin{figure}[t] 
\resizebox{68mm}{!}{\includegraphics[clip]{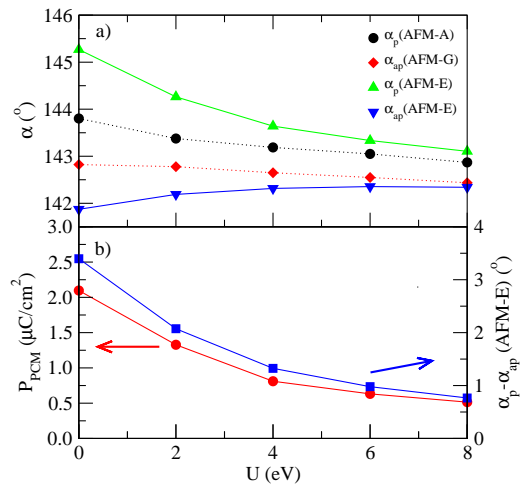}}
\caption{(Color online)
a) Mn-O-Mn angles (in degrees) in the parallel ($\alpha_p$)
  and antiparallel ($\alpha_{ap}$) Mn spin configurations vs. the Hubbard
  parameter U.
 b) Polarization calculated by the point charge model and $\alpha_{p}
  - \alpha_{ap}$ in the AFM-E phase. }
\label{fig:U}
\end{figure}

Due to the small size of the Ho ionic radius, \hmo\ 
shows~\cite{hmo1} a highly distorted 
perovskite structure with the Mn-O-Mn $ac$-plane angle $\alpha_0
\approx 144^\circ$. The AFM E-type spin-alignment is stabilized due to this
strong distortion of the perovskite structure, as was suggested by previous
model~\cite{dagotto,Ishihara}, and first-principles~\cite{slv} reports.
With the optimized atomic positions obtained after imposing the spin order,
we calculated the FE polarization using the point charge model (PCM) with
nominal charges. The calculated $P_{\text{PCM}}$ is shown in Fig.~\ref{fig:U}b 
as a function of the Hubbard parameter $U$. $P_{\text{PCM}} (U)$ is a decreasing function,
which is consistent with the general ideas of the model calculations~\cite{ivan}.
In that model, the polarization appears due to
the difference ($\alpha_{p} - \alpha_{ap}$) between the Mn-O-Mn
angles corresponding to the bonds with parallel ($\alpha_{p}$) and
antiparallel ($\alpha_{ap}$) spins, which is a consequence of the Hund's
coupling and virtual electronic hopping. In turn, $U$ characterizes
the energy penalty paid for adding an additional electron on a Mn
site. Thus, increasing $U$ makes the virtual electron hopping less
favorable, which reduces $\alpha_{p} - \alpha_{ap}$ and, ultimately,
$P$. Nevertheless, as shown in Fig.~\ref{fig:U}, the effect of even
a very large $U=8$~eV is to decrease $P$ by less than half an order of
magnitude. As the value of $U$ for \hmo\ is not known from experiments, 
we will resort to a parameter-free DF treatment with $U=0$ case from now onwards, 
noting that the calculated quantities should be trusted at least with respect to their
orders of magnitude.

A close look at the geometrically optimized structure in the AFM-E phase reveals a
complicated picture of displacements  shown in
Fig.~\ref{fig:struc}b. The Mn atoms displace by 0.04~\AA\ with
respect to the initial centrosymmetric $Pnma$ structure. The $a$ and
$b$ components of these displacements of different Mn atoms 
compensate each other, whereas the $c$ components add up to a net displacement of
$0.01$~\AA\ per Mn atom along the negative direction of the
$c$-axis. Similarly, the in-plane O atoms are displaced on average by 0.02 \AA\ 
with the resulting $c$-axis displacement of 0.01~\AA\ per atom in the
positive direction. Taking into account relatively smaller $c$-axis displacements
of interplane O and Ho atoms we obtain $P_{\text{PCM}}=2.1\, \mu$C/cm$^2$.
Therefore, our calculations with no fitting parameters independently
confirm the large electric polarization obtained in the previous
model~\cite{ivan}. In this regard, we hope that the experimentally attained
$P$ could be improved substantially as it was accomplished 
for another promising multiferroic BiFeO$_3$~\cite{neaton,review2}. 

To have a better understanding of the structural distortions caused by the magnetic
order, we also optimize the structural parameters for the
paraelectric AFM-A (all spins in the $ac$-plane are parallel) and AFM-G
(all neighboring spins are antiparallel) phases and compare them to the AFM-E
structure (see Table~\ref{tabella}). In agreement with the above
discussion, the calculated $\alpha_p$ for AFM-A is larger then
$\alpha_{ap}$ for AFM-G. However, when both types of bonds (parallel
and antiparallel) are present in the AFM-E phase, this difference is
even more pronounced, which explains the relatively high $P$. In addition, the
$U$-dependence of the angles is shown in Fig.~\ref{fig:U}a. 

\begin{table}
\caption{Relevant structural parameters for the AFM-A, G and E spin configurations: Mn-O-Mn angle
(in degrees) for parallel ($\alpha_p$) and anti-parallel
($\alpha_{ap}$) Mn spins, large ($d_l^1$ and $d_l^2$) and
small ($d_s^1$ and $d_s^2$) Mn-O bond-lengths (in \AA).}
\begin{tabular}{|c|c|c|c|c|c|c|}\hline \hline
& $\alpha_p$ & $\alpha_{ap}$ & $d_l^1$ & $d_l^2$ & $d_s^1$ & $d_s^2$ \\ \hline \hline
AFM-A & 143.8 & - & 2.20 & 2.20 & 1.93 & 1.93 \\ \hline
AFM-G & - & 142.8 & 2.24 & 2.24 & 1.90 & 1.90 \\ \hline
AFM-E & 145.3 & 141.9 & 2.25 &2.18 &1.92&  1.92 \\ \hline
\end{tabular}
\label{tabella}
\end{table}

\begin{figure}[t] 
\includegraphics[clip,width=78mm]{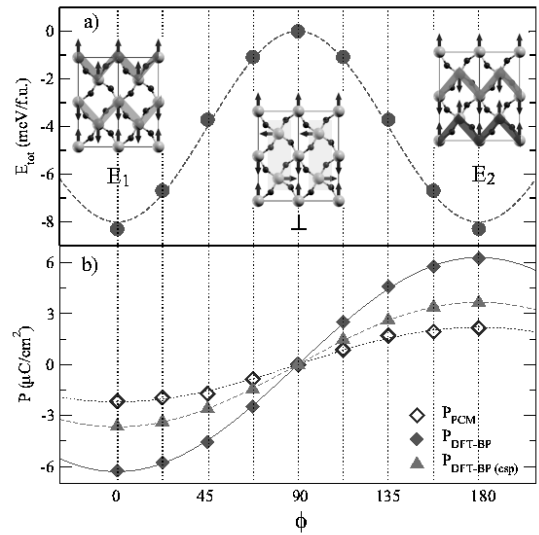}
\caption{(Color online)
a) Total energy as a function of the orientation of the
  central spins (see gray highlighted regions in the central $\perp$
  spin-configuration) with respect to the spin of the Mn in the
  origin. 
b) Polarization calculated via the PCM (empty blue diamonds) 
 and quantum-mechanically via DFT-BP (filled red diamonds). 
 The pink triangle denotes the values obtained via the DFT-BP approach
 for centrosymmetric atomic configurations as explained in the text. 
The lines are fits to $P \propto -\cos \phi$ with constant coefficients.} 
\label{fig:switch}
\end{figure}

The AFM-E phase shows two different kinds of AFM
domains~\cite{ivan}, $E_1$ and $E_2$ (see the left and right insets of
Fig. \ref{fig:switch}a, respectively), which are expected to show
opposite polarization, $-P_c$ and $P_c$.
In our calculational unit cell, $E_1$ and $E_2$
differ in the orientation of half of the spins (see grey highlighted
regions in the central inset).
Here, we consider  a FE-AFM switching path from
$-P_c$ ($E_1$) to $P_c$ ($E_2$) via a progressive rotation of the
central spins. 
According to the basic displacement-like mechanism for polarization,
we expect $P$ to switch from negative (in $E_1$) to positive (in
$E_2$) and to vanish when the relative orientation of the central
spins with respect to the fixed spins is close to
90$^\circ$. The 90$^\circ$ spin-configuration denoted as $\perp$ is
shown in the central inset of Fig. \ref{fig:switch}a. More precisely,
$\perp$ is an example of a spiral magnetic structure  similar
to that in TbMnO$_3$, but commensurate with the modulation vector $\mb
k = (1/2,0,0)$, and which should be FE
with $\mb P_\perp$ along the $c$-axis~\cite{nagaosa,Sergienko06,Mostovoy06,tbmno31}.
Based on the macroscopic symmetry considerations~\cite{supplinfo}, the
components of the polarization vector $\mb P$ can be expressed as
follows, 
\begin{equation}
\label{pcomps}
P_c = \chi_z (c_{xz}\sin \phi - c_{0} \cos\phi), \quad
P_a = c'_{xz} \chi_x \sin \phi, \quad P_b = 0
\end{equation}
where $\chi_x$ and $\chi_z$ are the components of the dielectric
susceptibility along the $a$ and $c$-axes, respectively, and 
$\phi$ is the rotation angle of the central spins. The coefficient $c_0$ stems from
nonrelativistic interactions, while $c_{xz}$ and $c'_{xz}$ originate
from the coupling of $P$ to the product of the $a$ and $c$ components of
the Mn spins, which has a relativistic origin. 

Equations~(\ref{pcomps}) lead to several important conclusions. First, for
the commensurate spiral state $\perp$ ($\phi=\pi/2$), a longitudinal component $P_a$
of the uniform polarization is present in addition to $P_c$. Second, $P_\perp$ is
finite due to purely relativistic effects, in agreement with previous 
microscopic models~\cite{nagaosa,Sergienko06}. Third, since the relatively small
relativistic effects such as SOC are neglected in our
computations, we observe that only the $c$ component
of the calculated $\mb P$ is finite for all $\phi$ and $P \propto \cos\phi$, in
excellent agreement with the numerical results in Fig.~\ref{fig:switch}b, which are discussed
below. Also, in our computations $P_\perp = 0$, and $\perp$ is taken as
the reference paraelectric structure with centrosymmetric positions (csp).

If FE switching is to occur, as the spin rotation proceeds
starting from $E_1$, the total energy is expected to increase up to a 
maximum corresponding to the ``paraelectric'' state ($\perp$), and then to
decrease again until the $E_2$ minimum is reached. Indeed, this 
happens when we perform the non-collinear calculations by varying $\phi$
between 0 and 180$^\circ$ with the full optimizations of the internal atomic
coordinates for each spin configuration. The calculated total energy
of the system as a function of $\phi$ (see Fig.\ref{fig:switch} a)
clearly shows a double-well structure, with the depth of the well of
about 8 meV/formula unit (f.u.). Although the exact magnitude of the
depth of the well can be affected 
by computational details and approximations, we expect the feasibility
of the magnetoelectric switching by the application of realistic
electric fields. The calculated energy barrier is, in fact,
smaller than in proper 
FE BaTiO$_3$ (18 meV/f.u.) and PbTiO$_3$ (200 meV/f.u.), and
multiferroic Ba$M$F$_4$ ($ > $ 20 meV/f.u.)~\cite{Cohen92,comp-other-fe}.

The evaluation of $P$ deserves a careful discussion, since it leads to an
intriguing outcome.
In Fig.\ref{fig:switch}b, we report the polarization evaluated by
the PCM and Berry-phase (BP) approaches within the
density-functional-theory (DFT-BP) along the previously mentioned
switching path. 
There is a marked disagreement between the PCM and DFT-BP
approaches, therefore suggesting that purely quantum electronic effects are at
play in determining the final $P$, similar to the 
conventional FEs~\cite{Vand94}. 
To investigate the purely electronic effects, 
we calculate$P$ considering the atomic positions of the
$\perp$ structure and artificially switching the 
spin-configuration, without relaxing the lattice degrees of freedom.
In this case, due to structural centrosymmetry, there is no
contribution from atomic displacements. However, the calculated BP
polarization is found to be up to 3.5 $\mu$C/cm$^2$ (see 
triangles in Fig.~\ref{fig:switch}b). This large contribution arises
solely from the electronic contribution due to symmetry breaking 
by the AFM-E ordering. 

\begin{figure}[htbp]
{\includegraphics[clip,width=70mm,angle=90]{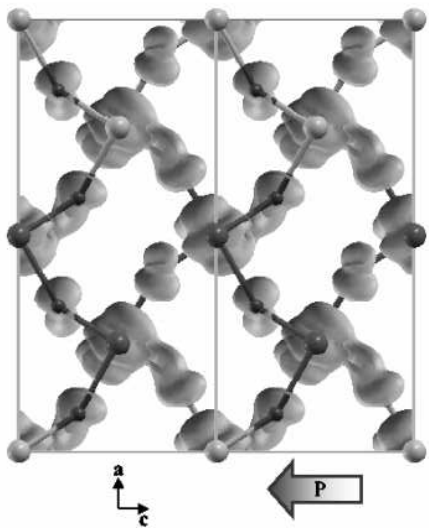}}
\caption{(Color online) The $ac$-plane charge density isosurface plot in the energy range [-0.8:0]
  eV (considering the top of the valence band as zero of the energy
  scale) for fully relaxed positions in the AFM-E1 spin
  configuration. Blue (green) balls denote Mn atoms with up (down)
  spins. 
}  
\label{fig:chg}
\end{figure}

The quantum effects can be quantified further by
considering the deviations of the effective Born charges from their
nominal valencies. In the  $\perp$ case, when
each O is coordinated to two Mn atoms with perpendicular spins, we obtain
$Z^*_\perp($Mn$) = 3.9\, e$ and $Z^*_\perp($O$) = -3.1\, e$.
Along the AFM-FE switching path, the O ions O$^p$ and O$^{ap}$ become 
increasingly different and acquire different Born charges due to the
orientation of the Mn spins to which they are bonded.
In the extreme points corresponding to
AFM-E1 and AFM-E2, we obtain: $Z^*($Mn$) = 3.8\, e$,
$Z^*($O$^{a})= -2.6 \,e$ and $Z^*($O$^{ap}) = -3.5\, e$.
Note, that in both $\perp$ and E-type spin arrangements, the Born
charges are not extremely different from
their nominal valence, consistent with a rather ionic nature of the
chemical bonds. We note that the two different kinds of O ions are responsible
for ferroelectricity in the case of centrosymmetric
positions, where the displacement mechanism is switched off. 
When atomic relaxation is allowed, all the atoms (including displaced
Mn) contribute to the final FE polarization. 
The inequivalence of O$^{a}$ and O$^{ap}$ is further confirmed by the
charge density plot for the fully-relaxed structure, corresponding to the energy
range of hybridized Mn $e_g$ and O $p$ orbitals located just below
the valence band maximum, as shown in Fig.~\ref{fig:chg}. 
In addition to the expected G-like orbital
ordering~\cite{Munoz01,dagotto}, Fig.~\ref{fig:chg} 
clearly demonstrates that there is a strong asymmetry in the
charge distribution between the two O ions. Moreover, focusing on
O$^{ap}$, the charge seems to favor the short Mn-O bond compared to the
long bond. This suggests that the polar charge distribution and
related wave-functions are due to a delicate combination of the 
Jahn-Teller effect and symmetry-breaking magnetic ordering.

In summary, our first-principles results show that, in the
AFM-E-type  \hmo, the symmetry-induced inequivalence of
the in-plane Mn-O-Mn configurations for parallel and antiparallel
spins is an efficient mechanism in driving a considerable
ferroelectric polarization. The calculated
total polarization of the AFM-E phase, $P \approx 6 \mu$C/cm$^2$ is
consistent with the previously obtained theoretical
estimates~\cite{ivan}. In addition to the displacement
mechanism, we find a larger but comparable contribution arising from a
purely electronic quantum effect of orbital polarization.
The finite ferroelectric polarization, even with a
centrosymmetric atomic arrangement, is an unambiguous indication of a
magnetism-induced electronic mechanism at play. 
Also, a magnetoelectric domain switching path is proposed, in which the reversal of
polarity of the applied electric field induces a 180$^\circ$-flip of 
selected spins. Although we have focused on the case of \hmo\ as example,
we believe our results concerning the dual nature of
ferroelectricity as arising from a symmetry breaking induced by the
magnetic order should have a wider validity for improper magnetic
ferroelectrics. 
Our findings suggest that the interpretation of experiments, as well as 
model calculations, should take into account $both$ the lattice and electronic 
mechanisms of improper ferroelectricity in multiferroics.

Computational support from Barcelona Supercomputing Center and 
Swedish National Infrastructure for Computing (SNIC) is acknowledged. 
Research at ORNL is sponsored by the Division
of Materials Sciences and Engineering, Office of Basic
Energy Sciences, U.S. Department of Energy, under Contract
No. DE-AC05-00OR22725 with Oak Ridge National
Laboratory, managed and operated by UT-Battelle, LLC. I.S. and E.D.
are supported in part by NSF Grant No. DMR-0443144.

\newpage
\newpage

\begin{widetext}

\begin{center}
\textbf{Auxiliary Material}
\end{center}

Here we give a brief derivation of selected results of the Landau
theory theory of phase transitions\cite{Landau} applied to the AFM-E order and its
coupling to ferroelectricity. Here we work in the $Pnma$ (\#62) space group
setting, which is related to $Pbnm$ used in Ref.~\onlinecite{ivan} 
via a different choice of the orthorhombic axes\cite{IT}. We also
place the origin in the Mn position. The order parameter of the
magnetic ordering ($\mb E_1, \mb E_2$) is written as a linear 
combination of Mn spins $\mb S_i$, where $i = 1,\ldots, 8$ denotes the
Mn atoms occupying the following positions in the magnetic unit cell,
corresponding to the modulation vector $\mb k = (1/2, 0, 0)$:
$$
1: (000),\quad 2: (\frac 1 2 0 \frac 1 2), \quad 3: (100),\quad 4:
(\frac 3 2 0 \frac 1 2), \quad 5: (0 \frac 12 0),\quad 6: (\frac 1 2 \frac 1
2 \frac 1 2), \quad 7: (1\frac 1 2 0),\quad 8: (\frac 3 2 \frac 1 2 \frac 1 2),
$$
The order parameter is obtained as\cite{ivan}
\begin{eqnarray}
\label{Ecomb}
\mb E_1 &=& \mb S_1 - \mb S_2 - \mb S_3 + \mb S_4 - \mb S_5 + \mb S_6 + \mb S_7 - \mb S_8,\nn\\
\mb E_2 &=& \mb S_1 + \mb S_2 - \mb S_3 - \mb S_4 - \mb S_5 - \mb S_6 + \mb S_7 + \mb S_8.
\end{eqnarray}
The components of $\mb E_1$ and  $\mb E_2$ transform according to the
rules given in Table~\ref{Rep}, where the transformation rules for electric 
polarization $\mb P$ are also summarized. Here, we analyze the
possible magnetoelectric coupling terms of the form $P\cdot  E^2$ in the
thermodynamic potential:
\begin{eqnarray}
\label{fme}
F_\text{me} &=& c_0 P_z(\mb E_1^{2} -  \mb E_2^{2}) + c_{yy} P_z(E_1^{y2}
- E_2^{y2}) + c_{zz} P_z(E_1^{z2} - E_2^{z2}) 
+ c_{xz} P_z(E_1^{x}E_2^{z}-E_2^{x}E_1^{z}) + c'_{xz} P_x (E_2^{z}E_2^{x}-E_1^{z}E_1^{x})\nn\\
&& + d_{xy} P_y(E_1^{y}E_2^{x}-E_2^{y}E_1^{x}) + d_{zy} P_y(E_1^{y}E_1^{z}-E_2^{y}E_2^{z}).
\end{eqnarray}
Clearly, only the first term 
has a nonrelativistic origin since $(\mb E_1^{2} -  \mb E_2^{2})$ is a 
linear combination of the terms $(\mb S_i \cdot \mb S_k)$.

Considering Eq.~(\ref{fme}) together with the usual dielectric energy $F_\text{de}
= 1/2 (\chi_x^{-1} P_x^2 + \chi_y^{-1} P_y^2 + \chi_z^{-1} P_z^2)$, we
obtain that $P_y = 0$ if $E_1^y=E_2^y=0$, which always holds in our
case. The other two components of $\mb P$ are given as
\begin{equation}
\label{pols}
P_x = - \chi_x c'_{xz} (E_2^{z}E_2^{x}-E_1^{z}E_1^{x}),
\quad 
P_z = - \chi_z [c_0 (\mb E_1^{2} -  \mb E_2^{2}) + c_{zz} (E_1^{z2} - E_2^{z2}) + c_{xz}
(E_1^{x}E_2^{z}-E_2^{x}E_1^{z})]. 
\end{equation}

As is proposed in the main text, the switching of the magnetoelectric
domain is accomplished by rotating the spins $\mb S_2$, $\mb S_4$,
$\mb S_6$, and $\mb S_8$ in the $xz$-plane. Thus, we set
$$
\mb S_1 = -\mb S_3 = -\mb S_5 = \mb S_7 = S (1, 0, 0),
\quad -\mb S_2 = \mb S_4 = \mb S_6 = -\mb S_8 = S (\cos \phi, 0, \sin \phi),
$$
Using Eqs.~(\ref{Ecomb}) and~(\ref{pols}), it is easy to verify that $\phi=0$ and
$\phi=\pi$ correspond to the domains $E_1^x (-P_z)$ and $E_2^x (+P_z)$,
respectively. Furthermore, we obtain,
$$
P_x = \chi_x 32 S^2 c'_{xz} \sin \phi, \quad
P_z = \chi_z (32 S^2 c_{xz} \sin \phi - 64 S^2 c_0 \cos \phi),
$$
which are equivalent to the expressions given in the main text with the
numerical constants and $S^2$ adsorbed in the coefficients.

As a by-product of the present analysis we also find that the
following term is allowed in the thermodynamic potential:
$$
F_\text{L} = \frac {\lambda}{V} \int (E_1^x \partial_x E_2^x - E_2^x
\partial_x E_1^x) dV.
$$
This \emph{Lifshitz}~\cite{Landau,Toledano} term leads to the following
typical series of phase transition when the temperature is lowered:
Para-phase $\rightarrow$ Incommensurate phase $\rightarrow$
``Locked-in'' commensurate phase\cite{Toledano}, which was indeed
observed in \hmo\cite{Munoz01}. It can be shown\cite{Toledano} that both $E_1^x$ 
and $E_2^x$ are modulated with $\mb k(T) = (1/2 + \delta k(T),0,0)$
in the AFM-Incommensurate phase. Taking into account the
magnetoelectric interactions~(\ref{fme}), this leads to a vanishing
macroscopic polarization in the incommensurate phase in agreement with
the experiment~\cite{Lorenz06}.

\begin{table}
\caption{\label{Rep}Matrices of the generators of space group $Pnma$ in the representations
spanned by $\mb E_1$, $\mb E_2$, and $\mb P$. The space group elements
are denoted $(r|hkl)$, where $r$ is the identity operation $1$,
two-fold rotation $2_{a,c}$, inversion $I$, or time reversal $1'$
followed by the translation $\boldsymbol \tau = h\mb a + k\mb b + l \mb c$.}
\begin{ruledtabular}
  \begin{tabular}{cccccc}
    & $(2_z|\frac 1 2 0 \frac 1 2 )$ & $(2_y|0 \frac 1 2 0)$ &
    $(I|000)$ & $(1|010)$ & $(1'|000)$\\
    \hline
    $\begin{matrix}
      E_1^x \\ E_2^x 
    \end{matrix}$ & 
    $ 
    \begin{matrix}
      1 & 0\\
      0 & -1
    \end{matrix}
    $ &
    $
    \begin{matrix}
      0 & 1\\
      1 & 0
    \end{matrix}
    $ &
    $
    \begin{matrix}
      0 & 1\\
      1 & 0
    \end{matrix}
    $ &
    $ 
    \begin{matrix}
      -1 & 0\\
      0 & -1
    \end{matrix}
    $ &
    $ 
    \begin{matrix}
      -1 & 0\\
      0 & -1
    \end{matrix}
    $\\
    \hline
    $\begin{matrix}
      E_1^y \\ E_2^y 
    \end{matrix}$ & 
    $ 
    \begin{matrix}
      1 & 0\\
      0 & -1
    \end{matrix}
    $ &
    $
    \begin{matrix}
      0 & -1\\
      -1 & 0
    \end{matrix}
    $ &
    $
    \begin{matrix}
      0 & 1\\
      1 & 0
    \end{matrix}
    $ &
    $ 
    \begin{matrix}
      -1 & 0\\
      0 & -1
    \end{matrix}
    $ &
    $ 
    \begin{matrix}
      -1 & 0\\
      0 & -1
    \end{matrix}
    $\\
    \hline
    $\begin{matrix}
      E_1^z \\ E_2^z 
    \end{matrix}$ & 
    $ 
    \begin{matrix}
     -1 & 0\\
      0 & 1
    \end{matrix}
    $ &
    $
    \begin{matrix}
      0 & 1\\
      1 & 0
    \end{matrix}
    $ &
    $
    \begin{matrix}
      0 & 1\\
      1 & 0
    \end{matrix}
    $ &
    $ 
    \begin{matrix}
      -1 & 0\\
      0 & -1
    \end{matrix}
    $ &
    $ 
    \begin{matrix}
      -1 & 0\\
      0 & -1
    \end{matrix}
    $\\
    \hline
    $P_x$ & -1 & -1 & -1 & 1 & 1\\
    \hline
    $P_y$ & -1 & 1 & -1 & 1 & 1\\
    \hline
    $P_z$ & 1 & -1 & -1 & 1 & 1\\
  \end{tabular}
\end{ruledtabular}
\end{table}

\end{widetext}

\end{document}